\documentstyle[aaspp4]{article}

\slugcomment{Scheduled for ApJ Letters, August 1, 1998 issue, Vol. 502}

\lefthead{Zampieri, Shapiro \& Colpi}
\righthead{Will a Black Hole Soon Emerge from SN 1997D ?}

\begin{document}

\title{Will a Black Hole Soon Emerge from SN 1997D ?}

\author{Luca Zampieri \altaffilmark{1},
Stuart L. Shapiro \altaffilmark{1,2} and
Monica Colpi \altaffilmark{3}}

\altaffiltext{1}{Department of Physics, Loomis Laboratory of Physics,
University of Illinois at Urbana--Champaign, 1110 West Green Street,
Urbana, IL 61801--3080}

\altaffiltext{2}{Department of Astronomy and National Center for
Supercomputing Applications, University of Illinois
at Urbana--Champaign, Urbana, IL 61801}

\altaffiltext{3}{Dipartimento di Fisica, Universit\`a degli Studi di
Milano, Via Celoria 16, I--20133 Milano, Italy}

\begin{abstract}

Observations combined with theoretical modeling of the light curve of 
the recently discovered supernova 1997D in NGC 1536 suggest that it might
host a black hole formed in the aftermath of the explosion.
We consider some observable consequences of a black hole in SN 1997D and
estimate the late--time accretion luminosity of the material which falls
back onto the hole. We find that this luminosity, decaying with  a 
characteristic power--law dependence on time, may emerge above the 
emission of the envelope in just a few years. Its detection would thus 
provide unmistakable evidence for the presence of a black hole.

\end{abstract}


\section{Introduction}


The nature of the compact remnant formed in the aftermath of a supernova
explosion depends on the mass of the iron core of the progenitor star,
on the amount of material that falls back onto the core
and on the maximum allowed neutron star mass ($M_{\rm {max}}$).
The core is usually in the 
range 1.4--2.0 $M_\odot$, while  the actual value of $M_{\rm{max}}$ is not 
conclusively established yet.
%
%
According to recent results on the properties of dense nucleon matter,
the maximum (nonrotating) neutron star mass falls in the
interval 1.8--2.2 $M_\odot$
(\markcite{Akmal}Akmal, Pandharipande \& Ravenhall 1998),
although smaller values
are not yet completely ruled out. Consequently, the amount of 
matter fallback plays a crucial role in determining the fate of the 
collapsed remnant. An investigation of the fallback process for 
stars with different masses and different explosion energies has 
been performed by \markcite{Woosley}Woosley \& Weaver (1995). They find 
that stars with 
main sequence mass larger than $\sim$ 25 $M_\odot$ have rather
massive cores ($\simeq 2 M_\odot$) and undergo accretion of a significant 
amount of matter, so that the remnant mass is larger than $3 M_\odot$, 
probably leading to the formation of a black hole (BH), after the launch
of a successful shock.
Fallback of a smaller amount of matter, $\sim 0.1-0.3 \,M_{\odot}$, may   
occur in explosions involving a progenitor of $19$--$25\, M_{\odot}$, 
although the actual value depends sensitively on the energy of the explosion.
Even in this case, a low mass ($\sim 1.5 M_{\odot}$) BH may form if
the equation for nuclear matter is not too stiff
(\markcite{Brown}Brown \& Bethe 1994).
The fact that a bright ($L \sim 10^{38}$ erg s$^{-1}$) accreting neutron 
star was not detected in SN 1987A in the early years following the 
explosion led to the hypothesis that a light underluminous BH may
have formed (\markcite{Chevalier}Chevalier 1989;
\markcite{Brown}Brown \& Bethe 1994; 
\markcite{Timmes}Woosley \& Timmes 1996).
Woosely \& Timmes (1996) first suggested that a
signature of BH formation in the light curve
would be a bright plateau that falls steeply to very low or zero
luminosity because of the low radioactive emission from heavy nuclei
present in the ejecta. Recently,
\markcite{Zampieri}Zampieri, Colpi, Shapiro \& Wasserman 
(1998a; hereafter ZCSW) pointed out that a BH 
may leave a distinguishable imprint in the light curve, which
could be revealed when the main sources of radioactive decay fade away
(see also \markcite{Zampierib}Zampieri, Colpi, Shapiro \& Wasserman 1998b).
Using a spherically symmetric, general--relativistic,
radiation--hydrodynamic Lagrangian code, \markcite{Zampieri}ZCSW showed that
fallback may lead to an accretion luminosity decaying secularly
with time as  $L\propto t^{-25/18}$.
For SN 1987A, the accretion luminosity turns out to be too dim 
to be detectable, at present, above the radioactive decay emission 
of $^{44}$Ti.

Recently, a new supernova, SN 1997D, was discovered in NGC 1536
(\markcite{DeMello}De Mello \& Benetti 1997).
The very low absolute luminosity and the low
inferred heavy element abundance in the ejecta
make this supernova a promising candidate for the detection
of a compact remnant. Following the suggestion by
\markcite{Turatto}Turatto {\it et al.\/} (1998), we consider
the consequences of a BH forming in SN 1997D. We give an
estimate of the accretion luminosity and the time of its emergence above the 
radioactive decay emission of the envelope.

\section{The peculiar Type II supernova 1997D} \label{s2}

\begin{figure}
\epsscale{0.4}
\plotone{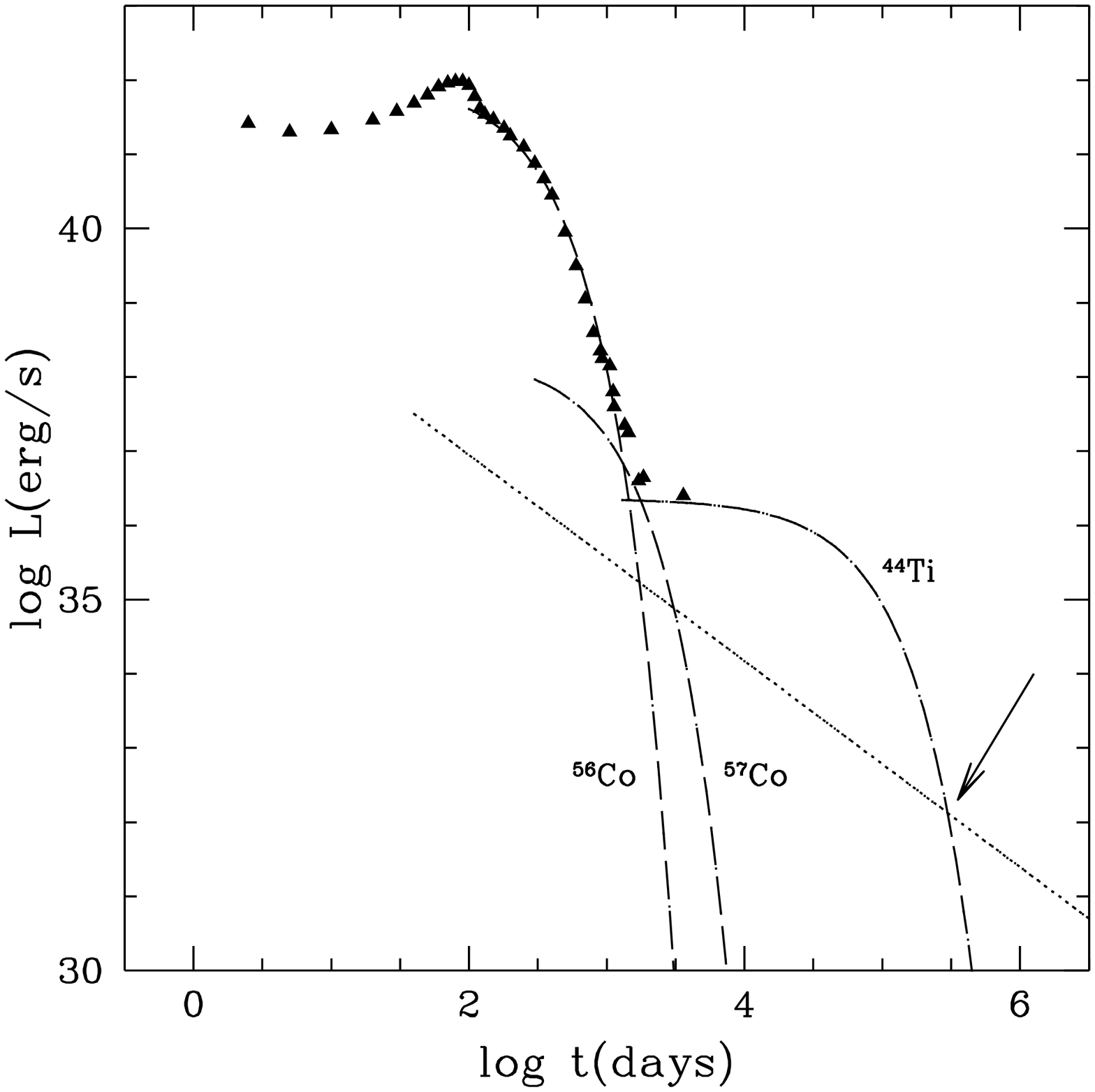}
\caption{Bolometric light curves for SN 1987A (from ZCSW).
The luminosity $L$ is plotted as a function of time $t$.
The {\it triangles} show the observed bolometric luminosity of SN 1987A.
The {\it dashed lines} represent the expected contribution from the 
decay of radioactive elements (0.07 $M_\odot$ of
$^{56}$Co and $5 \times 10^{-5} M_\odot$ of $^{57}$Co and $^{44}$Ti).
The {\it dotted line} denotes the calculated bolometric
luminosity emitted via accretion onto
a putative BH in SN 1987A. The {\it arrow}
marks the time of the BH emergence. \label{fig2}}
\end{figure}

\begin{figure}
\epsscale{0.4}
\plotone{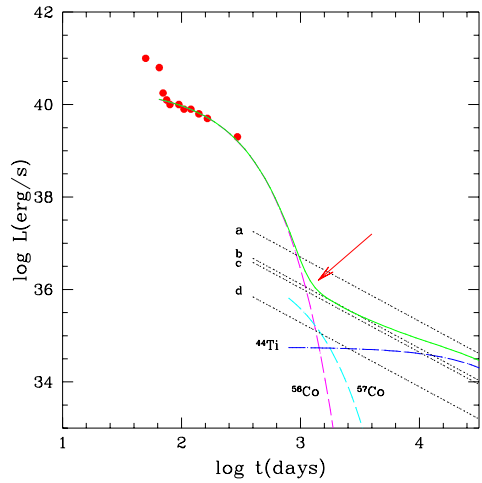}
\caption{Same as in Figure 1 for SN 1997D. The data for
SN 1997D {\it circles} are taken from Turatto {\it et al.\/} (1998).
The {\it  dashed lines} represent the expected contribution from the 
decay of radioactive elements (0.002 $M_\odot$ of
$^{56}$Co and $1.25 \times 10^{-6} M_\odot$ of $^{57}$Co and $^{44}$Ti).
Curves a, b, c and d denote the calculated
accretion luminosity at late times
for the following values of the initial density and expansion
timescale: a) $t_0 = 2$ hr, $\rho_0 = 0.1$ g cm$^{-3}$,
b) $t_0 = 2$ hr, $\rho_0 = 2 \times 10^{-2}$ g cm$^{-3}$,
c) $t_0 = 1$ hr, $\rho_0 = 0.1$ g cm$^{-3}$,
d) $t_0 = 2$ hr, $\rho_0 = 2 \times 10^{-3}$ g cm$^{-3}$.
The {\it continuous line} is the total calculated bolometric light curve 
for case c and the arrow marks the time of the BH emergence for this case.
\label{fig3}}
\end{figure}

SN 1997D was serendipitously discovered on January 14, 1997 by
\markcite{DeMello}De Mello \& Benetti (1997) during an observation of the 
parent galaxy NGC 1536.
Although the discovery of this supernova was made after maximum phase,
a number of closely spaced observations of NGC 1536 set an upper limit
to the optical luminosity $\sim 2$ magnitudes lower
than that of a typical Type II supernova. This result makes SN 1997D
possibly the least luminous Type II supernova ever observed.
A comparison of theoretical models with observations seems to indicate that
the epoch of the explosion probably occurred 50 days before the 
supernova was discovered (\markcite{Turatto}Turatto {\it et al.\/} 1998).
During all the evolutionary stages, the bolometric luminosity of SN 1997D
is orders of magnitude lower than that of the ``prototype'' SN 1987A
(see Figures \ref{fig2} and \ref{fig3}).
At around 70--80 days after the explosion,
the emission starts to be powered by $\gamma$--ray energy deposited
into the ejecta and, as shown also by SN 1987A, the decline of the tail 
of the light curve follows closely the decay rate of $^{56}$Co, the decay 
product of $^{56}$Ni.
If all the $\gamma$ rays are thermalized, the light curve after few weeks
is expected to follow the relation
\begin{equation}
L = L_0 \left( \frac{M_{\rm Co}}{M_\odot} \right) e^{-t/\tau_0} \, ,
\label{lumco}
\end{equation}
where $L_0 = 1.4 \times 10^{43}$ erg s$^{-1}$, $\tau_0 = 111$ days
is the decay time of $^{56}$Co and $M_{\rm Co}$ is the total mass of 
$^{56}$Co present in the ejecta.
Assuming for SN 1997D the same $\gamma$--ray deposition as in SN 1987A and
using the ratio of the luminosities in the tails of the light curves,
\markcite{Turatto}Turatto {\it et al.\/} (1998) estimate the mass of 
$^{56}$Ni ejected by SN 1997D to be about 40 times less than that in 
SN 1987A, i.e.
0.001--0.004 $M_\odot$. Another peculiar property of SN 1997D is
the low expansion velocity of the ejecta deduced from measurements
of the shift of the absorption lines in the spectrum. It turns
out to be $\sim 1200$ km s$^{-1}$.


Using a spherical, Newtonian, Lagrangian hydrodynamical code to simulate the 
explosion, \markcite{Turatto}Turatto {\it et al.\/} (1998) were able
to produce a satisfactory fit to the light curve of SN 1997D.
To fit the data,
a progenitor model with a radius $r = 2 \times 10^{13}$ cm and a
total mass $M = 26 M_\odot$ was adopted. The explosion energy was taken
to be $E = 4 \times 10^{50}$ ergs and the low luminosity tail from radioactive
decay is reproduced by the ejection of $M_{\rm{Ni}} = 0.002 M_\odot$ of 
$^{56}$Ni.
This figure is in agreement with that estimated by rescaling the value
of SN 1987A from the ratio of the luminosities of the tails.
The small values of $r$ and $E/M$ account for both
the observed low peak luminosity and expansion velocity.


The isotope $^{56}$Ni is produced in the deep layers of the star,
at the outer edge of the ``iron core''. In order to eject only a small amount
of $^{56}$Ni, a considerable fraction of the innermost stellar material
has to fall back onto the central compact core where it is
photodisintegrated before emitting $\gamma$--rays. The usual way to account
for this effect in hydrodynamic simulations of the explosion is to impose
a so--called ``mass cut'' in the post--explosion structure of the star.
In the $26 M_\odot$ model of SN 1997D of 
\markcite{Turatto}Turatto {\it et al.\/} (1998)
the mass cut has been placed at 1.8 $M_\odot$. This guarantees the ejection
of a mass of $^{56}$ Ni as low as 0.002 $M_\odot$.
The location of the mass cut implies that the mass of the
central compact remnant is $\ga 1.8 M_\odot$.
Then, as suggested by \markcite{Turatto}Turatto {\it et al.\/} (1998),
a BH could have formed 
and the low luminosity tail of SN 1997D
might be a signature of the formation of such an object.

In the following we will need an estimate of the
abundance of $^{57}$Co and $^{44}$Ti in the ejecta of SN 1997D.
As an upper limit to the abundance of these two isotopes,
we take the value inferred for SN 1987A
($\sim 5 \times 10^{-5} M_\odot$) and reduce it by a factor
of 40, similarly to what estimated for the abundance of $^{56}$Ni.

\section{Late--time accretion luminosity} \label{s3}

In a recent paper we have presented the first self--consistent, fully 
relativistic investigation of supernova fallback in presence of a BH
(\markcite{Zampieri}ZCSW).
We computed the light curve from a hot expanding
hydrogen ``cloud" in which the emitted energy comes from the release of
the heat residing in the original gas or generated by compression
in the course of accretion onto a central BH. Radioactive 
energy sources were not included in these calculations.
At the onset of evolution the cloud is optically thick
and has homogeneous density $\rho_0$; it is set into homologous expansion 
with a velocity profile $u=r/t_0$, with 
$t_0$ the expansion timescale.
The maximum velocity of the ejecta
is $V_0 = u(r_{out}) = r_{out}/t_0$, where $r_{out}$ is the outer radial
boundary.
During evolution, the unbound shells propagate outwards while those
with an outward velocity smaller than the escape speed 
fall back after turnaround, causing accretion on a secular timescale.
Initially, the cloud is 
radiation--dominated and photons diffuse outwards  
from the expanding stellar envelope. Hydrogen then recombines
and the ``cloud'' becomes optically thin. Depending on the
initial parameters, a significant amount of internal energy stored up into
the gas is liberated, giving rise to a bump in the light curve. The 
density and velocity profiles display a self--similar character and are
unaffected by the propagation of the recombination front.
After the emission maximum, the gas internal energy is nearly exhausted
and the luminosity falls off abruptly.

Soon after the recombination wave has propagated inward to the inner
accreting region, the front stalls because 
the compressional heating due to accretion surpasses radiative cooling.
From this moment on, the light curve is entirely powered by accretion:
the central BH ``becomes visible''.
The numerical calculations show that, at this stage, the accretion rate
${\dot M}$ in the inner regions is nearly independent of radius.
The evolution in the
%
luminosity--accretion rate plane closely follows the  analytic
curve $L=L({\dot M})$
derived by \markcite{Blondin}Blondin (1986)
for stationary optically thick hypercritical spherical accretion onto BHs:
\begin{eqnarray}
\frac{L}{L_{Edd}} & \simeq & 4 \times 10^{-7}
\left( \frac{\mu}{0.5} \right)^{-4/3}
\left( \frac{k_{es}}{0.4 \, {\rm cm^2 \, g^{-1}}} \right)^{-1/3} \times
\nonumber \\
& & 
\left( \frac{M_{bh}}{M_\odot} \right)^{-1/3}
\left( \frac{\dot M}{\dot M_{Edd}} \right)^{5/6} \, ,
\label{lumblon}
\end{eqnarray}
where $M_{bh}$ is the BH mass, $L_{Edd} = 4 \pi G M_{bh} c/k_{es} 
= 1.3 \times 10^{38} (k_{es}/0.4 \, {\rm cm^2 g^{-1}})^{-1} (M_{bh}/M_\odot)$
and ${\dot M}_{Edd} = L_{Edd}/c^2 
= 2.3 \times 10^{-9} (k_{es}/0.4 \, {\rm cm^2 g^{-1}})^{-1}
(M_{bh}/M_\odot) M_\odot$ yr$^{-1}$ are
the Eddington luminosity and accretion rate
and $k_{es}$ is the electron scattering opacity.

The late--time evolution of the accretion rate turns out to be in fair 
agreement with that of the hydrodynamic models
computed by 
\markcite{Colpi}Colpi, Shapiro \& Wasserman (1996; hereafter CSW)
for polytropic flows with no net radiation flux.
In fact, prior to recombination, the flow is radiation dominated and
very optically thick to emission--absorption so that, from
the dynamical point of view, it behaves like an adiabatic fluid gas
with $\Gamma = 4/3$.
On the other hand, by the time that recombination takes place and
the flow becomes transparent, radiation and pressure forces 
at around the accretion radius $r_a = GM_{bh}/c_s^2$ are 
very small and the hydrodynamic evolution turns out to be close
to that of a pressureless fluid (dust). Introducing the initial accretion 
timescale $t_{a,0} = GM_{bh}/c_{s,0}^3$ (where $c_{s,0}$ is the initial gas 
sound velocity) and the expansion timescale $t_0 = r_{out}/V_0$,
\markcite{Colpi}CSW have shown that, for envelopes with $t_{a,0}/t_0 < 1$,
pressure forces 
are important and ${\dot M}$ evolves along a sequence of quasi--static
Bondi--like states, 
with a slowly decreasing density at large distances. However, if the envelope
has $t_{a,0}/t_0 > 1$, then the gas behaves like
dust, and it can be shown that the late--time evolution of ${\dot M}$ 
is given by (\markcite{Colpi}CSW; equation [29])
\begin{equation}
\frac{\dot M}{\dot M_{Edd}} \simeq \frac{4\pi^{2/3}}{9} \rho_0 \,t_0 k_{es} c
\left ( \frac{t}{t_0} \right )^{-5/3} \, .
\label{dustmdot}
\end{equation}
Substituting equation (\ref{dustmdot}) into equation (\ref{lumblon}), we
obtain an analytic estimate of the late--time accretion luminosity
for a pressureless fluid (\markcite{Zampieri}ZCSW)
\begin{eqnarray}
&& \! \! \! \! \! \! \! \! L = 5.7 \times 10^{36}
\left( \frac{\mu}{0.5} \right)^{-4/3}
\left( \frac{k_{es}}{0.4 \, {\rm cm^2 \, g^{-1}}} \right)^{-1/2}
\left( \frac{M_{bh}}{M_\odot} \right)^{2/3} \nonumber \\
&& \left( \frac{\rho_0}{0.1 \, {\rm g \, cm^{-3}}} \right)^{5/6}
\left( \frac{t_0}{1 \, {\rm hr}} \right)^{20/9}
\left( \frac{t}{1 \, {\rm yr}} \right)^{-25/18}
\, {\rm erg \, s^{-1}} \, .
\label{lumas}
\end{eqnarray}

We will use equation (\ref{dustmdot}) to describe the 
evolution of the helium mantle of SN 1997D after the shock
has passed through the envelope and after the reverse shock (formed at the
hydrogen--helium interface) has reached the compact remnant (secondary 
fallback).

\section{Emergence of the black hole in SN 1997D}

The physical conditions of the He mantle after the passage of the shock, 
the amount of primary fallback and the mass of the central BH in
SN 1997D depend sensitively on the explosion energy.
Successful light curve and spectral fitting of SN 1997D show that
the energy of the explosion was small (a few $\times 10^{50}$ erg)
and a significant fraction of the He mantle
may have fallen back. Here we assume that, because of significant primary 
fallback, the mass of the BH is $M_{bh} \simeq 3 M_\odot$ (of which $\sim
1.4 M_\odot$ represents the mass of the iron core and $\sim 1.6 M_\odot$
is accreted from the helium mantle).
A few hours after the explosion, the internal energy in the He mantle of SN 
1997D turns out to be typically ${\tilde E} \simeq 10^{15}$ erg g$^{-1}$,
yielding a sound velocity $c_{s,0} = (\Gamma {\tilde E}/3)^{1/2} \simeq
2 \times 10^7$ cm $s^{-1}$ (where the polytropic index is $\Gamma = 5/3$ 
since the flow is gas--pressure dominated) and an accretion timescale $t_{a,0}
\simeq 3 \times 10^4$ s. The characteristic value of the
expansion timescale of the helium mantle inferred from low energy 
explosion numerical models turns out to be $t_0 \la$ 1--2 hr (T.R. Young;
private communication).
Since $t_{a,0} > \,t_0$, the late time hydrodynamic evolution is described 
within the dust approximation.
Accordingly, at late times the accretion luminosity of SN 1997D will be 
given by equation (\ref{lumas}) with $\mu = 4/3$ and $k_{es} = 0.2$ cm$^2$
g$^{-1}$ to account for the helium composition.

In Figures \ref{fig2} and \ref{fig3}, we plot the estimated late--time 
accretion luminosity of the He mantle of SN 1987A and SN 1997D.
For SN 1997D we plot four curves corresponding to different
initial values for the average density and the expansion timescale, which 
span a physically plausible range of the parameter space.
In order to bracket uncertainties in the value of the average 
density of the helium mantle a few hours after the explosion, we select 
three values of $\rho_0$ corresponding to outer radii in the interval
$r_{He} \simeq 3 \times 10^{11}$--$10^{12}$ cm.
As can be seen from Figure \ref{fig3}, {\it the putative BH in
SN 1997D should emerge above the radioactive decay emission of the envelope
approximately 3 years after the explosion}.
Since the luminosity by radioactive decay falls
off exponentially with time,
while the accretion luminosity drops only as a power law,
this estimate is not very sensitive to the actual value of the
initial parameters of the He mantle. An estimate of the
time $t_e$ of the BH emergence above the radioactive decay
emission of the envelope can be obtained by equating equations (\ref{lumco})
and (\ref{lumas}). We get
\begin{eqnarray}
\frac{t_e}{\tau_0} & - & \frac{25}{18} \ln \left( \frac{t_e}{\tau_0} 
\right) = 6.2 - \ln \left[ \left( \frac{\mu}{0.5} \right)^{-4/3}
\times \right. \nonumber \\
& 
& \left. \left( \frac{k_{es}}{0.4 \, {\rm cm^2 g^{-1}}} 
\right)^{-1/2} \left( \frac{M_{bh}}{M_\odot} \right)^{2/3}
\left( \frac{M_{\rm Co}}{10^{-3} M_\odot} \right)^{-1}
\times \right. \nonumber \\
&
& \left.
\left( \frac{\rho_0}{0.1 \, {\rm g cm^{-3}}} \right)^{5/6}
\left( \frac{t_0}{1 \, {\rm hr}} \right)^{20/9} \right] \, .
\label{temer}
\end{eqnarray}
Thus, $t_e \sim 6 \tau_0 \sim 2$ yrs,
and its value depends only logarithmically on 
the initial parameters.
On the other hand, as shown in Figure \ref{fig3},
in the four cases considered, the accretion luminosity at emergence 
(equation [\ref{lumas}]) varies significantly within the interval
$10^{35}$--$5 \times 10^{36}$ erg s$^{-1}$.
Thus, uncertainties in the inferred 
parameters of the mantle do not affect significantly the time of the 
emergence of the accretion luminosity above the 
emission of $^{56}$Co, but they do cause large uncertainties in the magnitude
of the luminosity at emergence, since $L \propto \rho_0^{5/6} t_0^{20/9}$.
Note that, for cases a, b, and c considered in Figure 2, the mass accreted
during secondary fallback is $\ga$ a few $M_\odot$, leading eventually 
to the formation of a rather massive black hole.

The possible presence of other radioactive isotopes in the ejecta adds an 
element of uncertainty to the above discussion.
As an upper limit in plotting the radioactive decay luminosity of $^{57}$Co
and $^{44}$Ti in Figure \ref{fig3} we have 
adopted the same rate of $\gamma$--ray deposition, at an abundance 40 times
lower, as in SN 1987A ($\sim 1.25 \times 10^{-6} M_\odot$; see Section 
\ref{s2} for a discussion).
As can be seen from the Figure, even in the least favorable case considered
(case d), the accretion luminosity at emergence is as large as
the decay luminosity of $^{57}$Co.

By comparing Figures \ref{fig2} and \ref{fig3}, it appears that SN 1997D 
is more promising than SN 1987A for detecting an accreting BH 
because of the lower amount of radioactive elements present in the envelope 
(which shifts the dashed curves downward) and the higher density of the 
He mantle (which moves the dotted curves upward).

\section{Discussion and conclusions}

We have shown that, if a BH has formed in SN 1997D, the
late--time accretion luminosity may dominate the bolometric light curve
$\sim$ 3 years after the explosion.
Depending on the actual value of the initial density and 
expansion timescale of the He mantle, the bolometric luminosity at 
emergence falls in the interval $L_e \simeq 2\times 10^{35}$--$10^{37}$ 
erg s$^{-1}$. After breakout, the light curve is expected to fall off 
as $t^{-25/18}$.
To estimate the apparent magnitude at emergence, we assume a black 
body emission spectrum and extrapolate from the present temperature to
get an effective temperature
$T_{eff} = 4,000$--$6,000$ K. This is reasonable since we expect that
the inner accreting region is ionized at $T \sim 10^4$ K and that dust 
may form in the outer ejecta reprocessing the radiation produced by
accretion and re--emitting it at a lower temperature.
Adopting a distance modulus $\mu = 30.64$ mag for NGC 1536,
the estimated apparent visual, red and I--band magnitudes at emergence are: 
$m_v \simeq 27$, $m_r \simeq m_I \simeq 26$ if $L_e \simeq 10^{37}$ erg 
s$^{-1}$
and $m_v \simeq 31$, $m_r \simeq m_I \simeq 30$ if $L_e \simeq 2 \times
10^{35}$ erg s$^{-1}$. If $L_e \ga 2 \times 10^{36}$ erg $s^{-1}$, the BH 
could be detectable for a few years in the red and I band with the Hubble 
Space Telescope, until the luminosity drops below threshold.
Clearly, the precise value of the initial parameters is crucial in
establishing the luminosity at emergence and hence whether or not the 
accreting BH can be detected.
Even if it is only marginally visible ($L_e \sim 2 \times 10^{36}$
erg $s^{-1}$), subtracting off the $^{56}$Co decay luminosity would reveal 
a power--law decay accretion luminosity which would confirm the presence 
of a BH.
We note that a supernova with properties similar to SN 1997D exploding 
in a nearer galaxy would appear much brighter and could reveal the 
presence of a newly formed BH from the distinct behavior of the 
late--time light curve.

The $t^{-25/18}$ power law decay of the light curve
is a signature of the presence of a black hole accreting
from the stellar envelope and follows directly from equations
(2) and (3). We note that this result is not valid for an accreting 
neutron star after the accretion shock has reached the trapping radius
(\markcite{Chevalier}Chevalier 1989). The reasons are that radiation 
pressure cannot be neglected at so early an epoch ($\la$ few years)
and the accretion luminosity efficiency 
is quite different.

A number of caveats apply to our result. The formation and the 
actual mass of the BH depend on the amount of mass which falls back
during the first hours after the explosion. 
This parameter is quite sensitive to the explosion energy, as shown by 
\markcite{Woosley}Woosley \& Weaver (1995). In addition, 
the state of the flow in the He mantle
is determined by the global hydrodynamics of the explosion.
%
%
The estimate of the late--time accretion luminosity given in Section
\ref{s3} assumes that the opacity in the accreting region is contributed
mainly by electron scattering and that the thermal
and dynamical effects of radioactive heating can be neglected.
Early in the evolution, in the inner accreting region, the temperature is
relatively high ($T \ga 10^7$ K) so that all elements are almost 
completely ionized. In these conditions, the dominant contribution to the 
Rosseland mean opacity comes from electron scattering and hence our
assumption is reasonable. Later on, the temperature decreases and
bound--free and bound--bound transitions (especially those of heavy elements)
may be important.
A lot of photons could remain trapped in the flow.
Analogous considerations apply to the expanding envelope at late 
times. For example, if the mean opacity of the envelope is $\sim 10^3 
k_{es}$, the time taken by photons to diffuse out of the
He mantle would be $\sim 20$ years for the typical
parameters of SN 1997D. Finally, in deriving our estimate we have
assumed that, after 1--2 years, the amount of heat deposited by
$\gamma$--rays from radioactive decay
is not sufficiently large to alter the character of the
flow. Given the small amount of heavy elements in the ejecta of
SN 1997D, this assumption seems reasonable.

Improved radiation--hydrodynamic calculations are necessary to provide a 
theoretical guide for identifying the nature of the compact remnant.

\acknowledgments

It is a pleasure to thank Massimo Turatto, Tim Young and their 
collaborators for providing us with data for the helium mantle of SN 
1997D in advance of publication and for several useful discussions.
We thank the referee for useful comments.
This work was supported in part by NSF grant AST 96--18524 and NASA 
grants NAG--5--2925 and NAG--5--3420 at the University of Illinois at 
Urbana--Champaign.

\end{document}